\documentclass[]{spie}  

\usepackage{amsmath,amsfonts,amssymb}
\usepackage{graphicx}
\usepackage[colorlinks=true, allcolors=blue]{hyperref}

\title{MYSTIC: Michigan Young STar Imager at CHARA}

\author[a]{John D. Monnier}
\author[a,b]{Jean-Baptiste Le Bouquin}
\author[c]{Narsireddy Anugu}
\author[c]{Stefan Kraus}
\author[a]{Benjamin R. Setterholm}
\author[a]{Jacob Ennis}
\author[b]{Cyprien Lanthermann}
\author[b]{Laurent Jocou}
\author[d]{Theo ten Brummelaar}

\affil[a]{University of Michigan, Ann Arbor, MI 48109 USA}
\affil[b]{Institut de Planetologie et d'Astrophysique de Grenoble, Grenoble, France}
\affil[c]{University of Exeter, Exeter, UK}
\affil[d]{CHARA Array, Georgia State University, Atlanta, GA}

\authorinfo{Further author information: (Send correspondence to J.D.M.)\\J.D.M.: E-mail: monnier@umich.edu, Telephone: 1 734 763 5822}

\begin{document} 
\maketitle

\begin{abstract}
We present the design for MYSTIC, the Michigan Young STar Imager at CHARA.  MYSTIC will be a K-band, cryogenic, 6-beam combiner for the Georgia State University CHARA telescope array.  The design follows the image-plane  combination scheme of the MIRC instrument where single-mode fibers bring starlight into a non-redundant fringe pattern to feed a spectrograph.  Beams will be injected in polarization-maintaining fibers outside the cryogenic dewar and then be transported through a vacuum feedthrough into the $\sim$220K cold volume where combination is achieved and the light is dispersed. We will use a C-RED One camera (First Light Imaging) based on the eAPD SAPHIRA detector to allow for near-photon-counting performance.  We also intend to support a 4-telescope mode using a leftover integrated optics component designed for the VLTI-GRAVITY experiment, allowing better sensitivity for the faintest targets.  Our primary science driver motivation is to image disks around young stars in order to better understand planet formation and how forming planets might influence disk structures.
\end{abstract}


\section{Introduction}
MYSTIC is an National Science Foundation (NSF, USA) funded project to build a 6-beam K-band combiner for the CHARA interferometer.  The instrument will be cryogenic and have two combiners. One combiner will be an image-plane design similar to the Michigan Infrared combiner (MIRC)\cite{mirc2004} while a second combiner will be based on the 4-beam integrated optics combiner GRAVITY\cite{gravity2010}.  The new instrument will use a C-RED One\cite{credone2016} camera (First Light Imaging) based on the SELEX-SAPHIRA\cite{finger2014} chip.  We expect ``first light'' in 2019.

The team designing and building MYSTIC presently consists of John Monnier (PI, U. Michigan), Jean-Baptiste Le Bouquin (UM, Institut de Plan\'etologie et d'Astrophysique de Grenoble) S. Kraus (Exeter), Narsireddy Anugu (Exeter), Benjamin Setterholm (UM), Jacob Ennis (UM), Cyprien Lanthermann (IPAG), Laurent Jocou (IPAG) and Theo ten Brummelaar (CHARA).  This development also benefits strongly from critical support of IPAG and CHARA on key issues.
There is strong software and hardware synergy with the MIRC-X instrument project led by Dr. Stefan Kraus (see elsewhere in these proceedings, see Kraus et al, Anugu et al.).  

The MYSTIC integrated optics combiner would not be possible without the technology developed by ESO and the GRAVITY Consortia. In order to develop strong  scientific synergy between the GRAVITY and MYSTIC projects, we have formed a preliminary ``MYSTIC-GRAVITY YSO Key Science Team'' for eventually coordinating a joint observing program focused on Young Stellar Objects. GRAVITY will be observing a large number of T Tauri and Herbig Ae/Be stars with exquisite precision for short baselines (B$<$120m) and MYSTIC will critically augment the Fourier coverage at longer baselines.  Combining our data on objects observable by both CHARA and VLTI will require careful coordination between our teams. Because GRAVITY and MYSTIC share the same combiner and detector architecture, we will also share technical knowledge and data reduction expertise.

The material presented in this paper was part of a ``Design Review'' in March 2018. We thank the review board for their practical advice and comments on the design, including some notable improvements that came out of the process. The review board consisted of Theo ten Brummelaar (CHARA), Judit Sturmann (CHARA), Mike Ireland (ANU), Denis Mourard (Nice), Laurent Jocou (IPAG), J.P. Berger (IPAG), and Antoine Merand (ESO, chair).

This paper will present the science motivation, hardware, and software architecture for MYSTIC.  The first section gives a brief overview of the top-level science requirements and priority observing modes.  Next, the ``Warm Optics'' will be described, including how MYSTIC and MIRC-X will work together at CHARA on the same optical bench. The third section will detail the optical design of the cryogenic ``Cold Optics,'' including the fibers, microptics, spectrograph and cryostat.   Detailed properties of the C-RED One camera will be described elsewhere (see Lanthermann et al., Anugu et al).   Lastly, the software architecture in common with MYSTIC and MIRC-X will be explained.

\section{Science Motivation}
\subsection{Imaging Young Stellar Objects}
The primary science goal of both MYSTIC and MIRC-X is to image the inner regions of the protoplanetary disks around Herbig Ae/Be and T Tauri stars with low spectral resolution, where MIRC-X will cover the J/H-band and MYSTIC the K-band. Earlier observations\cite{tannirkulam2008} have shown that these objects are very resolved on the long CHARA baselines, in many cases down to a visibility contrast $<=$10\%. To illustrate the regime we will likely be operating in, we selected three stars that cover typical cases, including a very resolved, bright Herbig Ae star (MWC275) a less resolved and fainter Herbig Ae star (like v1295 Aql) and one of the brightest T Tauri stars (SU Aur).  See Table~\ref{tab:science1}.

\begin{table}[ht]
\caption{Properties of typical Young Stellar Object targets for MYSTIC and MIRC-X} 
\label{tab:science1}
\begin{center}       
\begin{tabular}{|l|l|l|l|} 
\hline
\rule[-1ex]{0pt}{3.5ex} & MWC 275    & v1295 Aql & SU Aur  \\
 & (Herbig Ae) & (Herbig Be) & (T Tauri) \\
\hline
J/H/K uncorrelated & 6.2 / 5.5 /4.8 & 7.2 / 6.6 / 5.9 & 7.2 / 6.6 / 6.0 \\
\hline
Visibility on & 0.3? / 0.2 / 0.1 & 0.5? / 0.3 / 0.2 & 0.6? / 0.5? / 0.4? \\
longest baseline & & & \\
\hline
J/H/K correlated & 7.5 / 7.2 / 7.3 & 8.0 / 7.9 / 7.6 & 7.8 / 7.3 / 7.0 \\
\hline 
\end{tabular}
\end{center}
\end{table}

The computed correlated magnitudes illustrate that the long-wavelength bands are likely ideal for fringe detection and fringe tracking, i.e. H-band for MIRC-X and K-band for MYSTIC. However, there are likely also objects with higher J-band correlated flux, e.g. where the H-band is near a visibility null or the stellar flux contributions push the visibility up in the J-band. Therefore, we would like to optimisze the sensitivity for H-band, but be able to detect and track fringes in J-band as well.

Besides their low correlated flux in J/H/K-band, many stars in our YSO sample are also faint in R-band (down to R$=$13), which can result in poor tip-tilt performance and a poor beam quality. We hope the new CHARA adaptive optics system will improve performance in this regard (see contribution in these proceedings by ten Brummelaar et al.).

\subsection{Precision closure phases}
Another key science case is use precision closure phases for the direct detection of Hot Jupiters\cite{zhao2011}  and also to enable astrometric detection of planets in binary systems\cite{gardner2018}. This requires the measurement of high-precision closure phases\cite{monnier2007} with minimal baseline cross-talk. It also requires a minimum spectral resolution or R$\sim$300 in order to fill most of the entire primary lobe inside the interferometric field-of-view.  The GRAVITY combiner in particular is a pair-wise system with nearly zero cross-talk between baselines and may prove key to achieving these science goals.

\subsection{Spectro-interferometry}
Besides the low spectral resolution (LR) mode, we plan to implement also a medium spectral resolution (MR) mode for spectro-interferometric observations in spectral lines. 

\begin{figure} [ht]
   \begin{center}
   \begin{tabular}{c} 
   \includegraphics[width=6in]{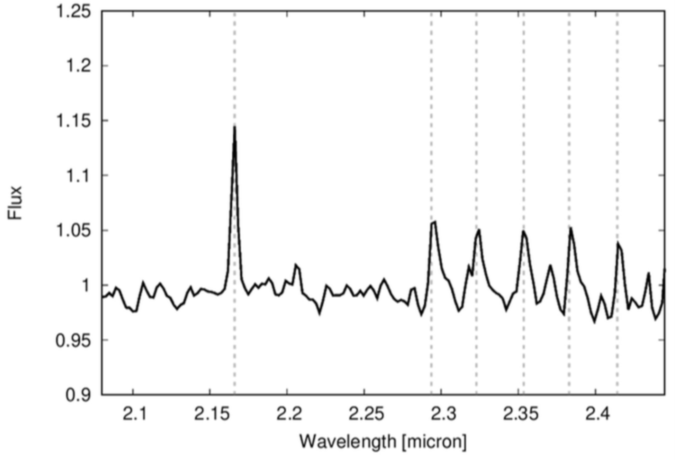}
   \end{tabular}
   \end{center}
   \caption[example] 
   { \label{fig:science2} 
Example of spectral lines that we plan to target with MYSTIC: K-band spectrum measured with GRAVITY on the massive YSO IRAS17216-3801, showing prominent Br-$\gamma$ and CO bandhead emission (Kraus et al\cite{kraus2017}).

}
   \end{figure}

The primary line tracers for MIRC-X are in the J-band, namely He-I 1.08$\mu$m, Pa-$\gamma$ 1.094$\mu$m, and Pa-$\beta$ 1.282$\mu$m, even though the Pa-$\beta$ line might not be accessible due to the 1.319$\mu$m metrology laser line that needs to be filtered out. There are also interesting line tracers in the H-band from the higher-transition Brackett series and some forbidden metallic lines. However, these lines appear prominently only in very few YSOs and are typically also weaker than the J-band transitions. For MYSTIC, the primary line tracers are the Br-$\gamma$ 2.166$\mu$m and CO bandhead 2.28-2.35$\mu$m lines in the K-band -- see Figure~\ref{fig:science2}.

Obtaining high-SNR in these line tracers requires long integration times in the science beam combiner (e.g. MIRC-X for He-I line observations) and puts strict requirements on the co-phasing performance of the instrument that acts as fringe tracker. Therefore, in the current plan we limit ourselves to a MR mode for velocity-integrated imaging of the line-emitting region (R$=$1170 for J-band; R$=$1035 for H-band; R$=$1700 for K-band). A high spectral resolution mode (R$>=$5000) could be implemented at a later time, once it has been demonstrated that the fringe tracking performance will be sufficient to enable integration times of a few seconds on the science combiner.

\subsection{Polarimetry}

In the intermediate term, we plan to commission a spectro-polarimetry mode to measure the interferometric observables along two parallel polarization directions and to derive Stokes Q and U parameters. The polarimetric mode will be used with the low spectral resolution prism (LR mode). We do not have sufficient resources for a full commissioning of this mode and will leave this for the future.

\subsection{Summary of priorities}

MYSTIC Priorities:
\begin{enumerate}
\item Sensitivity in K-band with at least four beams
\item Six-beam combination optimized for imaging 
\item Simultaneous operation with MIRC-X for wider instantaneous wavelength coverage and also to allow fringe tracking by one combiner in conjunction with high spectral resolution imaging with the other
\item High precision closure phases
\item Spectro-interferometry in Br-$\gamma$ 2.166 $\mu$m and CO bandhead 2.3$\mu$m lines
\item Polarimetry science mode (lower priority for MYSTIC than MIRC-X since scattering efficiency in K-band is much lower than in J/H-band)
\end{enumerate}

In order to fulfill these goals, we will have the following spectral resolutions available: R$=20$ (most sensitive mode, used with GRAVITY chip), R$=50$ (most sensitive imaging mode for 6 beams), R$=100$ (for increased field-of-view), R$=300$ (for wide binaries near limit of interferometric field-of-view), R$=900$ (to fully map field-of-view, R$=$1700 for Br-$\gamma$ and CO bandhead work when tracking with MIRC-X).

\section{Warm Optics}
\subsection{Overview}
For MYSTIC, the Warm Optics includes everything from the pick-up dichroics, which extract the light from the CHARA beams, to the injection of light inside the single mode fibers. It also includes the feedthroughs of these fibers into the cryostat.

   \begin{figure} [ht]
   \begin{center}
   \begin{tabular}{c} 
   \includegraphics[width=6in]{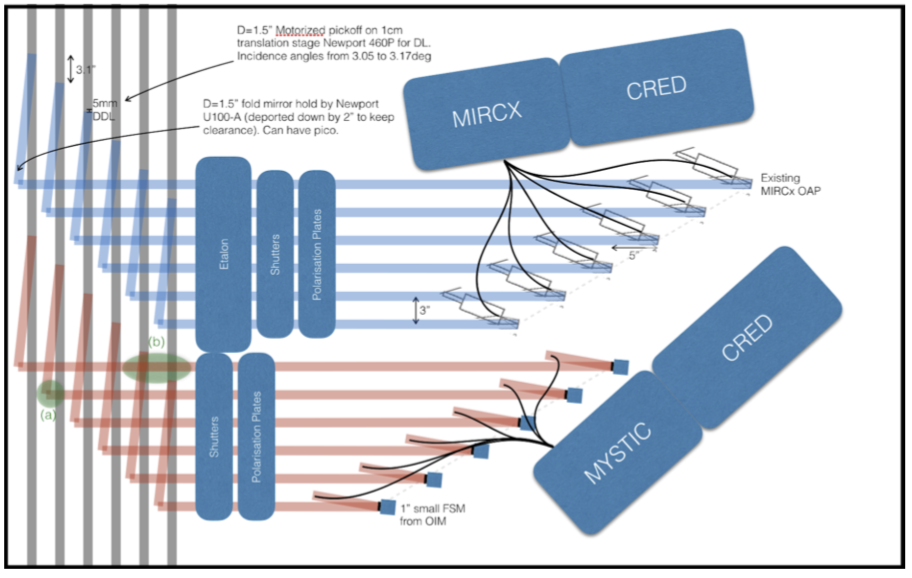}
   \end{tabular}
   \end{center}
   \caption[example] 
   { \label{fig:layout3} 
New layout of MYSTIC and MIRC-X beams and combiners using the existing 5'x8' optical table at CHARA.
}
   \end{figure}

\subsection{Table layout}
In order to make room on the existing MIRC optical table for MYSTIC, a re-design of the overall beam layout was needed.
Our new design for the table layout has been based on the following following goals, here listed from top priority to lower priority:
\begin{itemize}
\item For MIRC-X: use the existing fiber injection modules. For MYSTIC: develop new injection modules that separates image quality and tip/tilt correction.
\item Access the MIRC-X and MYSTIC beam combiners from the edge of the table.
\item Avoid moving the fibers when actuating optical path difference.
\item Use off-the-shelves components.
\item Keep symmetry between MIRC-X and MYSTIC.
\item Minimize the number of reflections.
\item Place the wavelength-calibration etalon (used for astrometry experiment) in the MIRC-X beams only.
\item Allow the automation of the pick-up devices to allow MIRC-X and MYSTIC to easily be put in- or out- of the CHARA infrared beam train.
\end{itemize}

Figure~\ref{fig:layout3} shows our new table layout for MYSTIC and MIRC-X.  We point out the following details:
\begin{itemize}
\item The beams (K-band in red, J+H band in blue) are extracted from the CHARA beams. Each dichroic/mirror is mounted on an individual motorized flip-style pick-up and a translation stage. The difference of OPD is 3'' between beams, so the incidence angle on each dichroic is slightly different.
\item The beams are folded into a regular grid of 3'' spacing and 5'' OPD delay with a 1.5'' flat mirror. The mount of this mirror is motorized with pico-motors for occasional realignment of the pupil shear.
\item The beams cross the polarization controller plates and the shutters. In the case of MIRC-X, they optionally cross the Etalon (wavelength reference).
\item The tip/tilt for MYSTIC is compensated by the reflection on the Field Stabilization Mirror (FSM). MIRC-X still employs movable fibers for beam injection as before.
\item The beams are focused by the off-axis parabolas into the fibers.
\item There is extra room for a future upgrade to implement phase-induced amplitude apodization\cite{piaa2014} to improve fiber coupling.
\end{itemize}

   \begin{figure} [ht]
   \begin{center}
   \begin{tabular}{c} 
   \includegraphics[width=6in]{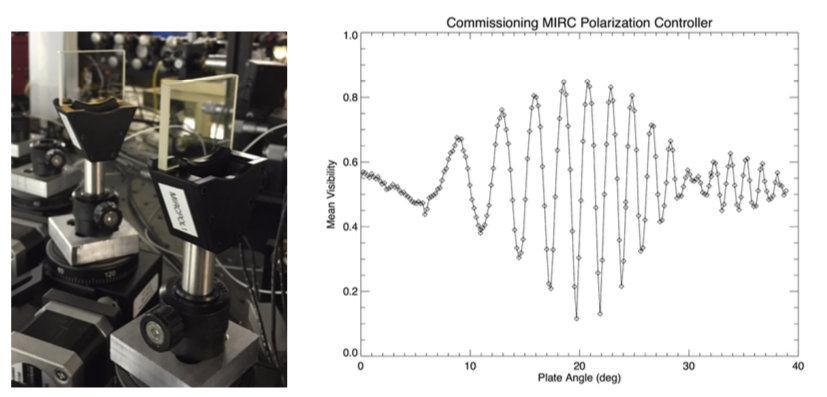}
   \end{tabular}
   \end{center}
   \caption[example] 
   { \label{fig:polarization4} 
Picture of the MIRC-X Polarization Plates (left) and wavelength-averaged fringe visibility versus the plate angle (right).  The same system will be used for MYSTIC but with different anti-reflection coatings.
}
\end{figure}

\subsection{Pickoff optic and differential OPD control}

We plan to motorize the pickoffs for MIRC-X and MYSTIC to allow improved integration with CHARA.  We will have 10mm of optical OPD adjustment in both MIRC-X and MYSTIC to align our combiners in-phase with other CHARA combiners.  Absolute accuracy will be within $<50\mu$m, in order to recover the coherence length after initialization.  We have kept the incidence angle low (3deg) to keep pupil shear below 0.75mm over the course of 5mm mechanical (10mm OPD).

For MIRC-X, we will use the motorized flip-up mounts 8892 from Newport equipped with flat mirrors. Six of them are already available. For MYSTIC, we will use  the dichroic of the former CHAMP\cite{champ2006} instrument, although sub-optimal performance at long-end of filter bandpass may require acquisition of new dicroics.  

The differential delay lines are planned to be Newport linear stages (460P) motorized with Zaber T-LA28A. Six of them are already available from CHAMP. It provides 20 mm-mechanical range with repeatability $<$4$\mu$m and step size of 0.1$\mu$m. The angular deviation of the mount in pitch/yaw is specified $<$150 $\mu$rad over the entire course.

We found that the repeatability of the flip-in is better than 100$\mu$rad and that the wobbling of the translation stage can be as large as 250$\mu$rad optical over several mm, in both axes. Fortunately, the wobbling over 100$\mu$m is less than 50$\mu$rad, meaning we do not expect to have to re-align tip-tilt for small OPD adjustments, e.g., to account for longitudinal differential dispersion between H and K bands.

\subsection{Polarization control}

The polarization controllers provide an adjustable phase-shift between the vertical/horizontal axes (birefringence axes of fibers) of up to 5 lambdas. This is ensured by rotation of thick Lithium Niobate plates following the approach of Lazareff et al\cite{lazareff2012}.

We will use 4~mm thickness, AR-coated z-cut LiNbO3 plates from Crylight. The company guarantee a transmission $>$98\% over the K-band for incidence angles from 0 to 30 deg, and a transmitted wavefront distortion $<$100nm. Two identical plates have been already tested in MIRC-X (see Figure~\ref{fig:polarization4}). This experiment demonstrated that our design can modulate more than 5 fringes around the mean angle 20deg, with a resolution better than lambda/20.  We have some concern how well this will work for MIRC-X when trying to phase up J and H band simultaneously, since there are some chromatic effects using this method of birefringence control.

MYSTIC will use the same motorized rotary stage developed and tested already for the MIRC-X experiment (AY110-60 from OES). It provides a resolution of 0.02deg, which corresponds to lambda/100 at 20 deg mean angle. The axis is controlled by the R256 from Lin Engineering with a CHARA--based controller and home-built homing switch.

   \begin{figure} [ht]
   \begin{center}
   \begin{tabular}{c} 
   \includegraphics[height=2in]{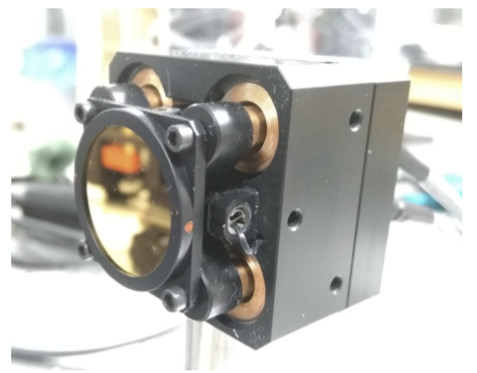}
      \includegraphics[height=2in]{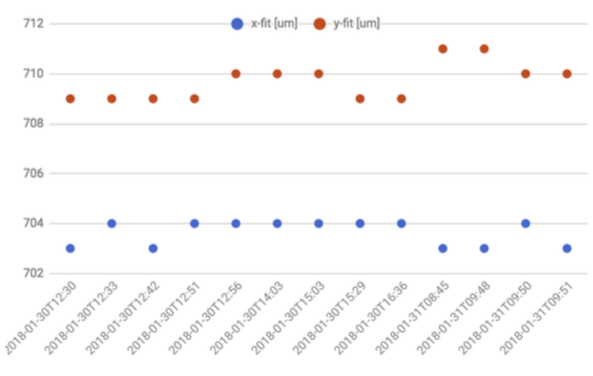}

   \end{tabular}
   \end{center}
   \caption[example] 
   { \label{fig:steering5} 
(left) The 1” field steering mirror (FMS) by Optics in Motion (OIM) being tested in the lab.  (right) Stability tests of FSM showing the x and y position of a reference spot versus time.  
}
\end{figure}

\subsection{Motorized tip-tilt}
The CHARA guiding axis is not strongly controlled to the MYSTIC/MIRC-X axis since instruments are on different tables and guiding is done with visible light without compensation for atmospheric refraction. Therefore, regular optimization of the tip/tilt remain critical for ensuring optimal injection into the fibers. We want to be able to re-do such ``fiber explorer'' as often as possible, and as fast as possible.

Our specifications for the tip-tilt control:
\begin{itemize}
\item Clear aperture: $>$22mm
\item Precision: 50 milliarcsecond-sky = 3arcsec-optic-lab  = 14 $\mu$RAD = 0.0006 deg
\item Range: $\pm$10arcsec-sky = $\pm$10arcmin-optic-lab = $\pm$2.8mRAD = $\pm$0.12 deg
\item Bandwidth: few Hz. But see above.
\item Encoders: yes. Closed-loop operation is mandatory.
\end{itemize}

We carried out market research to find appropriate actuators for our tip-tilt control. Open-loop pico-motors are not usable because they are not repeatable and have large backlash. Classical piezo solutions are very expensive ($>\$9$K each beam), and as such are not considered as long as a solution $<\$6$K is found. We also considered closed-loop picomotors from Newport, which are substantially more expensive than the open-loop versions.

The Fast Steering Mirror (FSM) of the Optics in Motion (OIM) company appeared to meet our specifications,  however the medium term stability ($\sim$hours) was not documented since these devices are primarily used for closed-loop tip-tilt systems.  Using a reference source, we monitored the position of best injection for several hours and found the mirror kept the spot position $\pm$0.5$\mu$m over several hours. This correspond to FWHM/4 for our PSF with MYSTIC.
 See Figure~\ref{fig:steering5} for a picture of the device and our test results.  Based on these tests, we have adopted the OIM FSM for our MYSTIC design.

\begin{table}[ht]
\caption{Properties of NUFERN PM1950 Fiber} 
\label{tab:fibers2}
\begin{center}       
\begin{tabular}{|l|l|} 
\hline
\rule[-1ex]{0pt}{3.5ex}Parameter & Value \\
\hline
Core NA & 0.2 \\
\hline
Cutoff & 1.72$\mu$m \\
\hline
Mode Field Diameter at 1.95$\mu$m & 8.0$\mu$m \\
\hline 
Core diameter & 7$\mu$m \\
\hline
Beat length at 1.95$\mu$m& 5.2mm \\
\hline
Transmission  & \\
 \qquad 1~meter @ 2.2$\mu$m & 94\% \\
 \qquad 1~meter @ 2.37$\mu$m & 85\% \\
\hline
\end{tabular}
\end{center}
\end{table}

   \begin{figure} [ht]
   \begin{center}
   \begin{tabular}{c} 
   \includegraphics[height=2.4in]{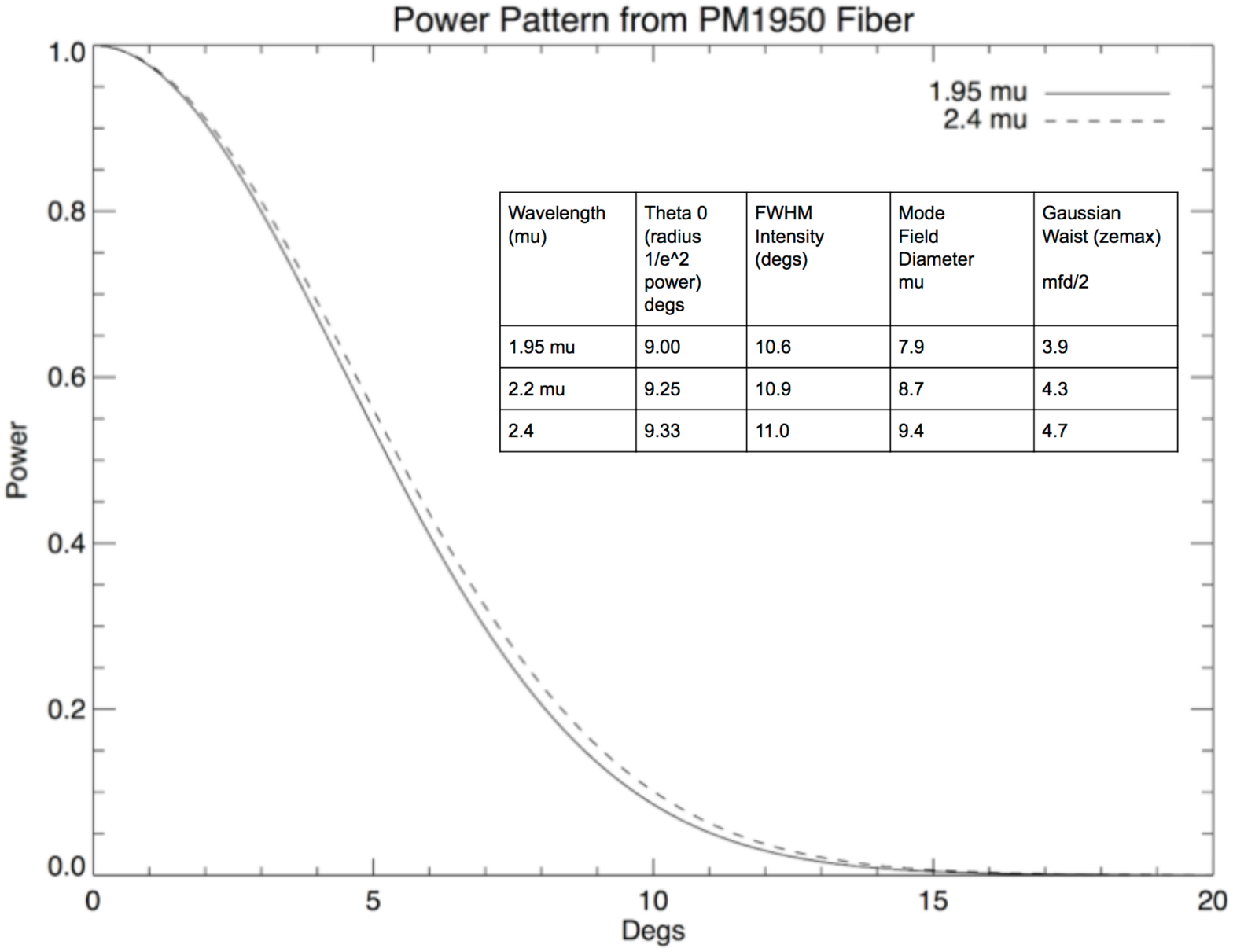}
      \includegraphics[height=2.2in]{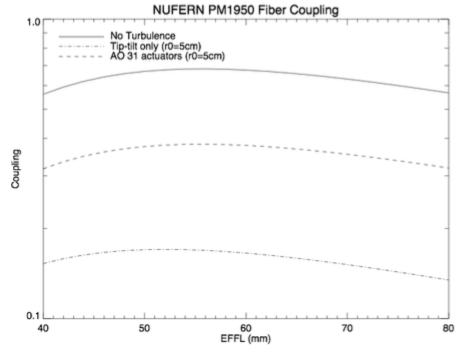}

   \end{tabular}
   \end{center}
   \caption[example] 
   { \label{fig:fiber6} 
(left) Calculated intensity pattern of K-band light exiting the Nufern PM1950 fiber. (right) Estimate of Coupling efficiency of light from a CHARA telescope into the NUFERN PM1950 fiber as a function of the focal length of the injection optics -- we calculated for perfect (diffraction-limited) conditions, for only tip-tilt control, and with the planned adaptive optics system. 
}
\end{figure}
\subsection{Fibers}
We chose off-the-shelf, silica-based specialty fibers for MYSTIC rather than the Fluoride fiber used in the GRAVITY combiner. This was mainly a cost issue and followed testing of candidate fibers to assure adequate throughput.

We obtained 50m of Nufern PM1950 with a cutoff wavelength of 1.72$\mu$m and NA 0.2, which closely matches the NA of the GRAVITY fibers (0.19).  Note that this is a polarization-maintaining single-mode fiber.  Over a fiber length of 1m, we measured 94\% transmission at 2.2$\mu$m and 85\% transmission at 2.37$\mu$m, with minimal bending losses. Table~\ref{tab:fibers2} contains a summary of the PM1950 properties.

   \begin{figure} [ht]
   \begin{center}
   \begin{tabular}{c} 
   \includegraphics[height=1.5in]{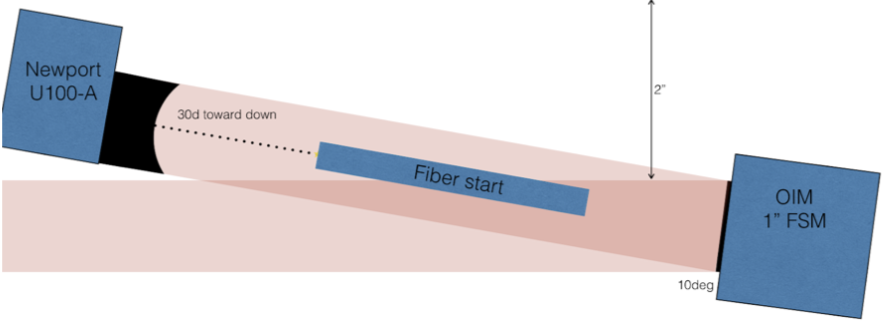}
      \includegraphics[height=1.5in]{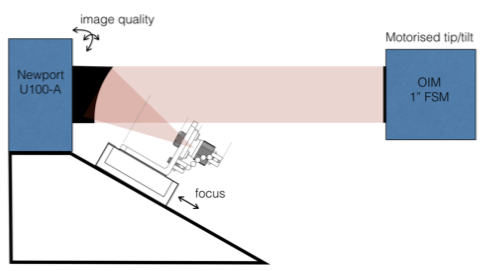}

   \end{tabular}
   \end{center}
   \caption[example] 
   { \label{fig:injection7} 
(left) Top view of the planned fiber injection module using a OIM FSM and an OTS Off-axis Parabola (e.g., Edmund Optics 1'' diameter, 54.4mm EFL and 30deg reflection with bare gold coating). (right) side view.  
}
\end{figure}

\subsection{Fiber injection module}

With a fiber selected, we can calculate the optimal focal length for the off-axis parabola we will use for injecting starlight into the fiber (the expected beam profile is shown in left panel of Figure~\ref{fig:fiber6}. The optimum f/D of the imaging system depends only on the mode-field radius $w$ (at which the intensity drops to 1/e$^2$) of the fiber, and is given by (no central obscuration): $F/D = 1.40 w  / \lambda$.
Considering our mode field diameter of $2w=8\mu$m, the optimal focal is F/D$\sim$2.87. As comparison and check, Shaklan \& Roddier\cite{shaklan1988} (1988) found an optimal F/D$\sim$5.2 for fibers with NA$=$0.11. Considering a simple proportionality and our NA$=$0.2, this corresponds to F/D$=$2.86, quite close to our value.  This corresponds to focal length of 54mm (see results from detailed numerical simulation in right pnael of Figure~\ref{fig:fiber6} for a 19mm beam with 5mm central obscuration with and without the CHARA adaptive optics system). Actually the range 48mm to 62mm provides efficient injection. The fastest side improves the robustness against turbulence by demagnifying the image motion, at the cost of 5\% injection loss.

Based on the above calculation, the product choices and the beam layout constraints, we show our injection module design in Figure~\ref{fig:injection7}.  Each MYSTIC beam will bounce off the OIM field steering mirror, hit the off-axis parabola (EFFL 54.4mm, Edmund Optics), and be reflected downward into the PM1950 fiber, connectorized with an FC connector.

   \begin{figure} [ht]
   \begin{center}
   \begin{tabular}{c} 
   \includegraphics[width=6in]{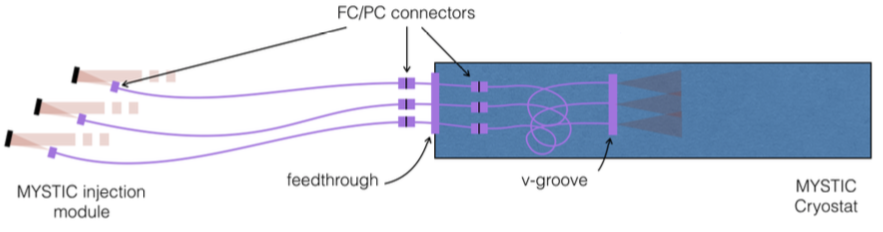}
   \end{tabular}
   \end{center}
   \caption[example] 
   { \label{fig:fiberlengths8} 
Sketch of the MYSTIC fiber link with 3 fiber sections. Only 3 out of the 6 beams are represented.  There is a section linking the off-axis parabola free-space injection to the fiber feedthrough via a FC/PC connector (``Link one''). Then there is a feedthrough between the air and vacuum spaces (``Link 2''). Finally there is a connection between the feedthrough and the fibers that go to the silicon v-groove arrays (``Link three'').
}
\end{figure}

\subsection{Fiber Equalization}

While MIRC-X has only one fiber link for each beam, MYSTIC will have 3 separate links in order to bring light into the cryostat. We need to separate into multiple links to allow for two combiner choices (Classic and Integrated Optics) and to reduce risk on both the cryostat feedthrough and silicon v-groove preparations. In addition, this choice allows us to insert a laser into Warm Optics or the Cold Optics without manipulating the fiber on the off-axis-parabola, by plugging it at the additional connection. The change of beam combiner will also be done at this additional connection, consequently two set of long warm section are required.

Figure~\ref{fig:fiberlengths8} shows the current plan.
The fiber lengths will be:
\begin{itemize}
\item Link one from off-axis parabola to feedthrough. One set with kength $\sim$0.6m
\item Link two is the feedthrough that connect the OAP fiber to the combiner fibers in the dewar.  Two sets with length $\sim$0.2m.
\item Link three from feedthrough to each combiner. Two sets with length 0.3m.
\end{itemize}

We numerically computed the contrast for broad K-band, and for ten spectral channels (R$\sim$50), for varying lengths of fiber. We conclude that an equalization better than 2~mm is required to keep contrast $>$0.95 inside each of the spectral channel in low resolution. This is a non-trivial specification for manufacturers.

We intend to have each link be internally matched to $\pm$2mm, that is, all the links from OAP to feedthrough will be matched, all the feedthroughs will be matched, and all the fibers attached the v-groove will be matched.  By mixing and matching links, we expect to meet our specification.  

If we find our length matching fails or we have birefringence mismatch, the backup solution is to insert glass plates into the beams to compensate for the dispersion. The glass plates can be of different thicknesses and be tilted for final tuning. The glasses will also introduce up to 2 mm of OPD, but this can be handled by our long internal OPD. Moreover, it is probable that this OPD will actually partially compensate the differential length of the fibers.

\section{Cold Optics}
\subsection{Overview} 

The Cold Optics of MYSTIC is composed of four subsystems: the Classic 6-beam combiner, the Integrated Optics combiner, the spectrograph and the camera.

   \begin{figure} [ht]
   \begin{center}
   \begin{tabular}{c} 
   \includegraphics[width=6in]{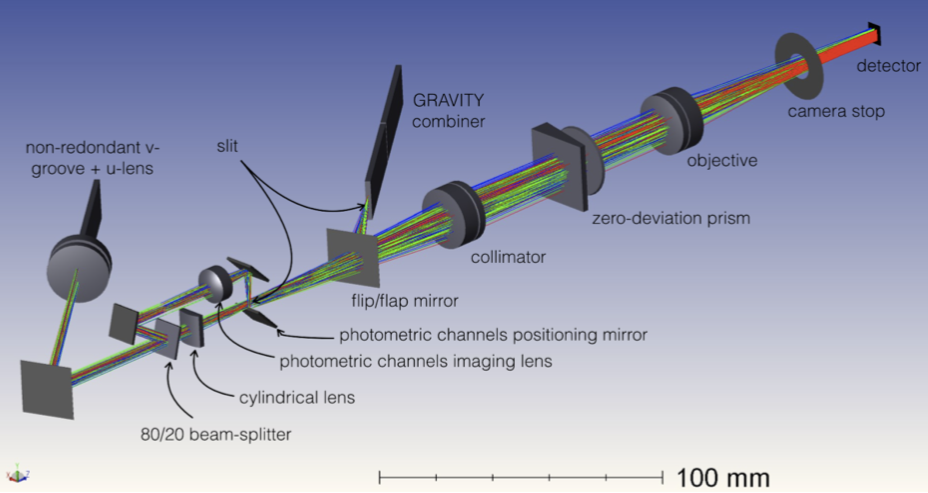}
   \end{tabular}
   \end{center}
   \caption[example] 
   { \label{fig:coldoverview9} 
Sketch of the MYSTIC cold optics including the classic combiner with its photometric channels, the Integrated Optics (GRAVITY) combiner, the spectrograph and the detector
}
\end{figure}

\subsection{``Classic'' combiner} 
The Classic combiner of MYSTIC is based on the MIRC design: it is a 6-beam all-in-one multiaxial (image plane) combiner. This fringe pattern is compressed in the spectral direction in order to define the input slit of the spectrograph.

The fringes are formed by overlapping the six diffracting beams originating from the non-redundant v-groove in slots 4, 6, 13, 18, 24 and 28 with pitch 250$\mu$m. This non-redundant pattern has been selected based on the following criteria:
Space of 2 fibers for baseline at shortest frequency to minimize the cross-talk with the frequency of the fringe envelope (also called DC spike).
Allow removal of 1 fiber to make a ‘cross-talk’ resistant design, where all baseline frequencies are separated by 2x fiber spacing.
Minimum total length, to minimize number of pixels we need to readout on the chip.

\begin{figure} [ht]
   \begin{center}
   \begin{tabular}{c} 
   \includegraphics[width=6in]{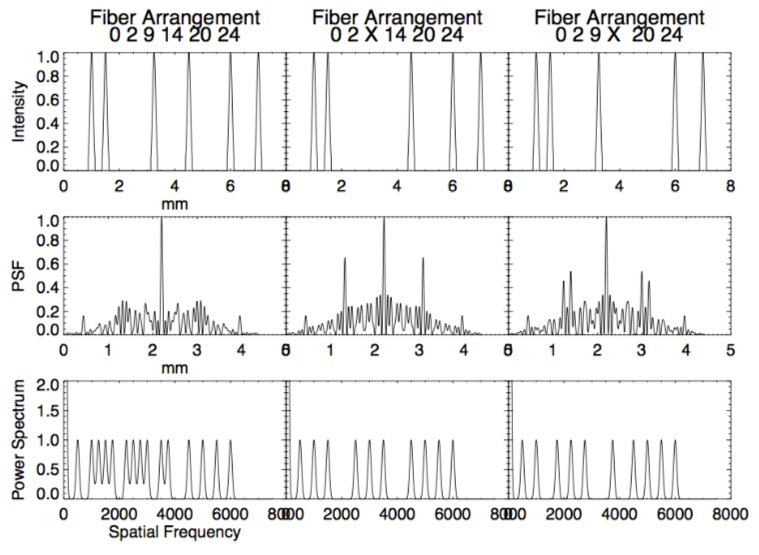}
   \end{tabular}
   \end{center}
   \caption[example] 
   { \label{fig:nrm10} 
Fiber arrangement for Classic combiner, fringes, and the resulting power spectrum. The middle and right are the two ‘cross-talk’ resistant 5-beam combination
}
\end{figure}

The maximum baselines for our design is 24 units, which means we have to read out 20\% more pixels than in the former MIRC design (which had longest baseline of 20 units) that did not have a ‘cross-talk’ resistant mode. Interestingly, the ‘cross-talk’ resistant mode is obtained by removing either fiber 3 or 4. By putting S1/S2 in these fibers, we will not lose much uv coverage by switching between a S1+E1E2W1W2 and then S2+E1E2W1W2, if the cross-talk resistant design is useful. See Figure~\ref{fig:nrm10} to see the fiber spacings for the 6-beam system and the two 5-beam cross-talk resistant patterns.  Also, included here is the the point-source function and the power spectrum for each pattern.

\begin{figure} [ht]
   \begin{center}
   \begin{tabular}{c} 
   \includegraphics[width=6in]{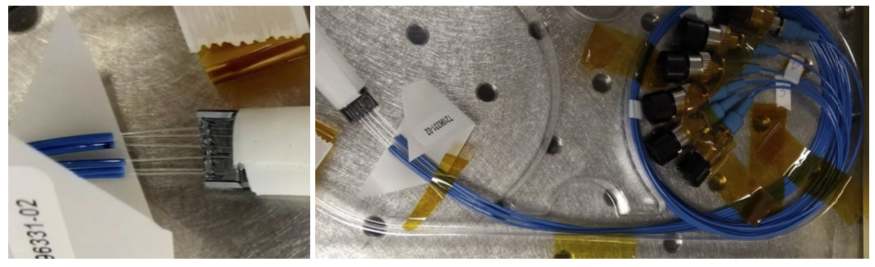}
   \end{tabular}
   \end{center}
   \caption[example] 
   { \label{fig:vgroove11} 
V-groove from OZ Optics connectorized by Coastal Connections.
This is the J/H-band version for MIRC-X with 1m fibers.
}
\end{figure}

OZ Optics placed the fibers in the silicon v-groove array and Coastal Connections equalized the fibers and connectorized the fibers for the first unit to be used in MIRC-X. Figure~\ref{fig:vgroove11} shows the actual hardware currently in Michigan.

\begin{figure} [ht]
   \begin{center}
   \begin{tabular}{c} 
   \includegraphics[height=1in]{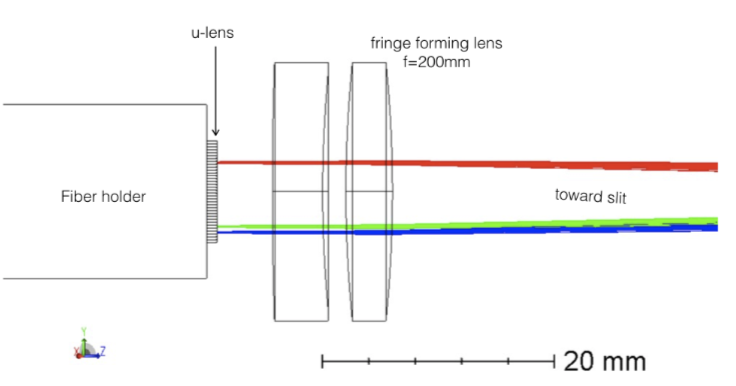}
   \includegraphics[height=1.5in]{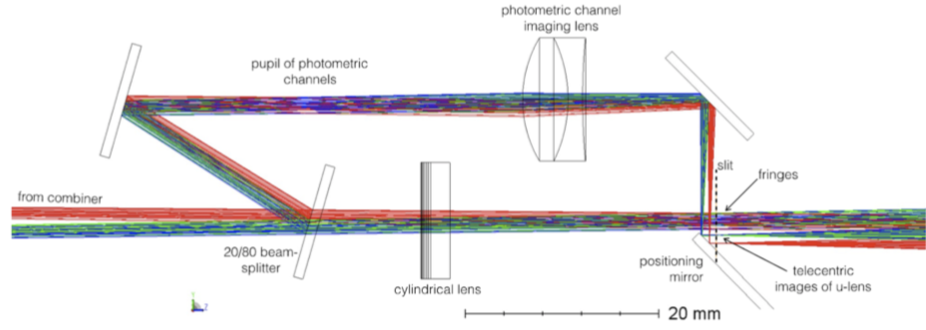}
   \end{tabular}
   \end{center}
   \caption[example] 
   { \label{fig:classic13} 
Optical train for classical image-plane mode.  Note the photometric channels are formed optically using a beamsplitter and focusing lens that projects the fiber images onto the same plane as the fringes at the pseudo-slit.  This is a different design than the current CHARA/MIRC design that used a set of fibers to capture and re-image the photometric channels.
}
\end{figure}

Once the beams leave the v-groove and are collimated by the micro-lens array, the diffracting beams are overlapped with a long focal lens (f$=$200 mm, Thorlabs achromatic doublet ACA254-200-D, optimized for 2.2$\mu$m). This provides a sampling of 2.75pix/fringes at wavelength 1.9$\mu$m considering the C-RED One pixel size of 24$\mu$m. 
We intend to design a custom microlens array that we will glue to the silicon v-groove array to improve stability of alignment inside the cryostat.   See Figure~\ref{fig:classic13} (left panel) to see the layout of the v-groove, microlens array and the fringe forming lens.

After the fringe-forming lens, there  are slit-forming optics to (1) compress the fringe pattern in the spectral direction to define the input slit of the spectrograph ; (2) image the beams individually into this slit in order to monitor their photometry. In the spectral direction (e.g orthogonal to the fringes) the overlapping beams are brought to a focus by the means of a cylindrical lens. This focus defines the input slit of the spectrograph. The 30 mm focal length of this cylindrical optics matches the f-number defined for the spectrograph. 

The Classic combiner also needs to monitor the flux of each beam, with so-called ``photometric channels.'' In MIRC, this is achieved by splitting 20\% of the flux shortly after the lenslet array and re-injecting it into multi-mode fibers whose outputs are re-arranged on the slit. This design has the following difficulties:
The 45deg reflection on the 20/80 beam-splitter is highly polarized.
The injection into the multi-mode fibers is hard to align.
The K-band transmission is reduced by the additional 30 cm of fibers

Our new design re-images the output of the lenslet array onto the slit. We split away 20\% of the light with a weakly-polarizing beamsplitter, sending the light to pass through an imaging lens rather than a cylindrical lens. The photometric channel imaging lens could be something similar to the Thorlabs achromatic doublet AC127-030-C, which has a 30 mm focal length although not optimzed for 2$\mu$m operation.
Interestingly, the optical layout permits to relay the pupil such that the images of the photometric channels on the slit are telecentric. This is ensured by placing the ``pupil'' of the photometric channels at the focus of the imaging lens.

See Figure~\ref{fig:classic13} for the optical layout of both the fringe and photometry beam paths leading to input slit of the spectrograph.

\subsection{Integrated Optics Combiner}

   \begin{figure} [ht]
   \begin{center}
   \begin{tabular}{c} 
   \includegraphics[width=6in]{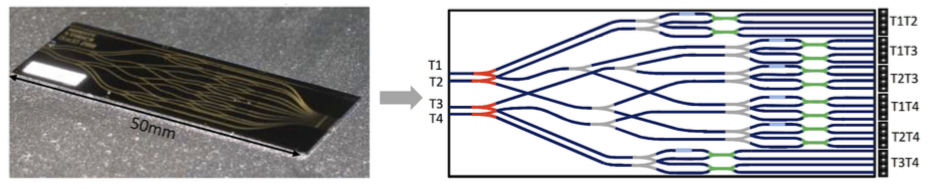}
   \end{tabular}
   \end{center}
   \caption[example] 
   { \label{fig:gravity14} 
Picture and sketch of the 4-beam pairwise GRAVITY combiner
}
\end{figure}

The integrated optics combiner chip is a leftover from the GRAVITY project. It implements a 4-beam pairwise ABCD combination. The chip has 24 single mode outputs at a pitch 180$\mu$m ($=$ 4.2 mm FOV) with a numerical aperture of NA$\sim$0.2 . This corresponds to a telecentric FOV of $\pm$2.1 mm. The outputs are separated by 7.5 pixels on the detector.  See Figure~\ref{fig:gravity14}.

One complication is that MYSTIC spectrograph operates at f/8 and not f/2 as for GRAVITY. We will glue a custom-designed microlens array to the end of the IO device to convert the light from f/2 to f/12. 

\begin{figure} [ht]
   \begin{center}
   \begin{tabular}{c} 
   \includegraphics[width=6in]{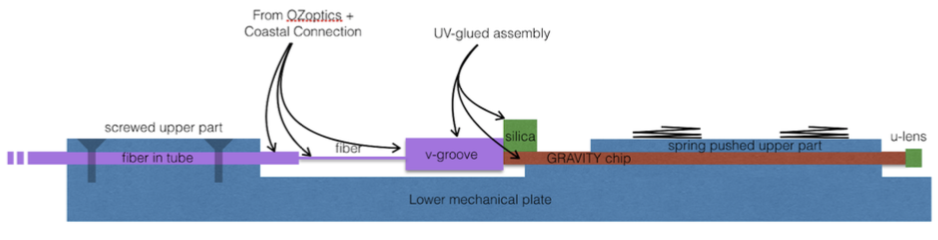}
   \end{tabular}
   \end{center}
   \caption[example] 
   { \label{fig:gravity15} 
This is a preliminary design for the assembly of the integrated optics combiner for MYSTIC based on the design use for VLTI/GRAVITY.
}
\end{figure}

Another challenge will be to glue the silicon v-groove to the IO device and to mount all the components in a robust way suitable for cryogenic operations and the resultant temperature variations.  This part of the development will be done at IPAG (Laurent Jocou) who has extensive experience in this arena.  See Figure~\ref{fig:gravity15} for a preliminary design sketch for the mounting.

\subsection{Spectrograph} 

\begin{figure} [ht]
   \begin{center}
   \begin{tabular}{c} 
   \includegraphics[width=6in]{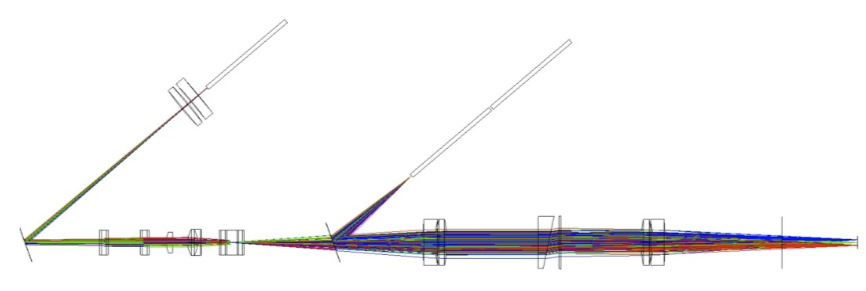}
   \end{tabular}
   \end{center}
   \caption[example] 
   { \label{fig:topview16} 
Top view of the overall cold optic
(colors are for wavelengths)
}
\end{figure}

The spectrograph is composed of a collimator, a spectral  and/or polarization dispersive elements (Wollaston), and an objective. The magnification from the slit to the detector is unitary.  An integrated view of both combiners with the spectrograph is shown in Figure~\ref{fig:topview16}.

\begin{figure} [ht]
   \begin{center}
   \begin{tabular}{c} 
   \includegraphics[width=6in]{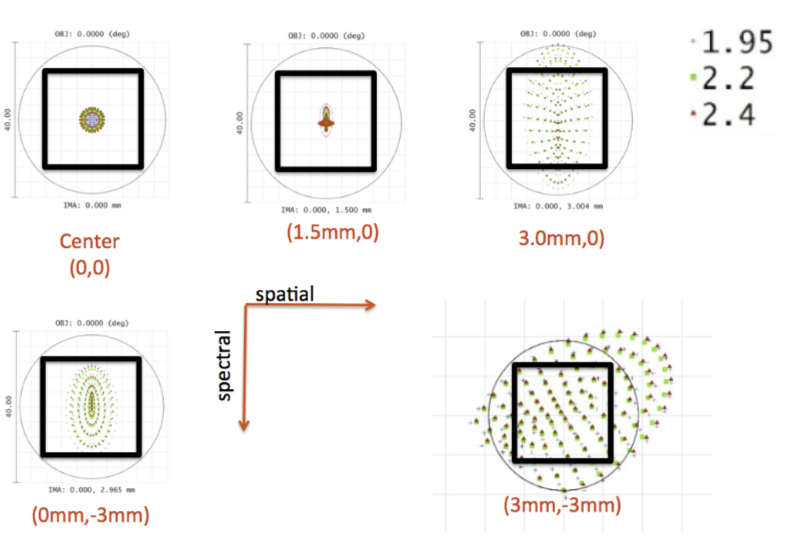}
   \end{tabular}
   \end{center}
   \caption[example] 
   { \label{fig:spectrograph17} 
Spot diagram for spectrograph assuming a f/8 uniform beam (Photometric channels of Classic combiner, output of GRAVITY combiner). 
The focus shifts by 300$\mu$m at T$=$210K.  The dark square box is the size of one pixel.
}
\end{figure}

The choice of the focal length and optics size is a compromise between the FOV on the collimator (shorter focal or bigger optics), and vignetting on the camera stop (longer focal). Thorlabs has 1” achromatic doublet, optimized for 2.3$\mu$m available in 75 mm, 100 mm, 150 mm and 200 mm (ACA254-xxx-D series). We selected the 100 mm for our design. The focal shift is 20$\mu$m from 1.9 to 2.4$\mu$m has low impact on image quality for our large f-number. The collimated beam size has a FWHM of 9.1 mm which intoduces only  small ($<5\%$) vignetting.  Figure~\ref{fig:spectrograph17} shows the spot diagrams for this spectrograph from our ZEMAX calculation.  The image quality is diffraction-limited for all of K band and for most of the field-of-view, except the far corner of the detector.

\subsection{Cryostat} 
The purpose of the cryostat is to ensure that all emissive surfaces seen by the camera are cold enough so that their thermal emission does not contribute to the final noise. The C-RED One camera is packaged inside a small standalone cryostat cooled by a pulsed tube cryocooler. The combiners and spectrograph will be packaged into a second cryostat.

The background flux depends on the camera quantum efficiency ($\sim$0.7), the transmission filters of the camera (cut-off at 2.45$\mu$m), the pixel size (24$\mu$m), the numerical aperture of the camera cold stop (f/4), and the integration time (25 ms). This corresponds to 5ph/pix at 230K, 1.5ph/pix at 220K and 0.7ph/pix at 215K. For this integration time, the camera noise is about 1.5e-/pix, dominated by the intrinsic dark current of the camera. Therefore the temperature of the combiner and the spectrograph should be 220K (goal 215K).

\begin{figure} [ht]
   \begin{center}
   \begin{tabular}{c} 
   \includegraphics[height=2in]{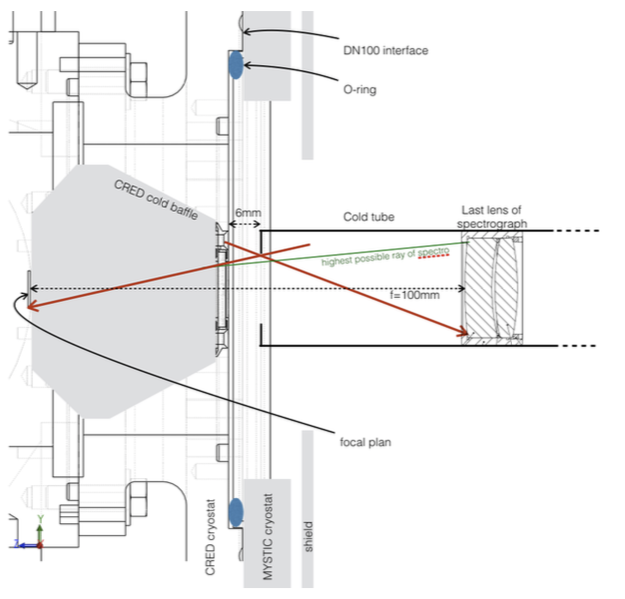}
      \includegraphics[height=2in]{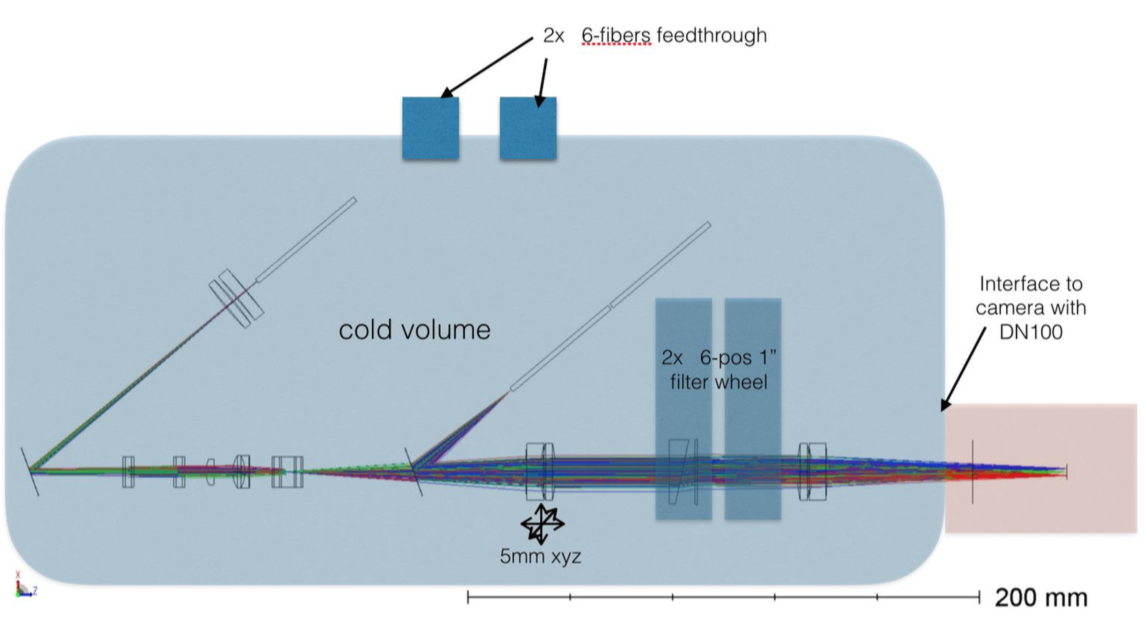}

   \end{tabular}
   \end{center}
   \caption[example] 
   { \label{fig:interface20} 
(left) Sketch of the interface between the C-RED One and MYSTIC cryostats. Rays indicate stray light paths that must be controlled for low background operations. (right)    Preliminary plan for placement of optics and feedthrough in the MYSTIC cryostat. 
}
\end{figure}

The MYSTIC cryostat will directly connected to the C-RED One cryostat with the available DN100 interface. The spectrograph includes a cold tube of 1'' that extends up to few mm from the cold baffle of C-RED One. It ensures that:
the camera ‘sees’ cold surfaces only, in order to avoid thermal background;
the last optics ‘sees’ cold surfaces only, in order to avoid any parasitic reflection of thermal background in the coated surfaces.  Figure~\ref{fig:interface20} (left) shows a detail drawing of the mechanical interface between the C-RED One and the MYSTIC cryostat, including the critical rays that need controlled for low-background operation.  We anticipate leaving the window in place during operation to separate the vacuum chambers of the two dewars, although we will have option to remove window if excessive background is encountered.

While we have not finalized our cryostat design, Figure~\ref{fig:interface20} shows one proposed geometry to fit the entire combiner inside 400mm x 200m x 200mm cold volume.

\section{Software}
\subsection{Overview}
This section will give an overview of the software environment for MYSTIC.  Our software plan has evolved from the DAQ code and GUI system used by the existing MIRC instrument.   Since summer 2017, the MIRC instrument runs with the new C-RED One camera.  Furthermore, a large effort is underway since 2017 to migrate the software into a more CHARA compliant architecture, based on client/server and GTK GUIs. 

The software architecture of MIRC-X and MYSTIC are in common, based on the same high speed camera system (C-RED ONE).  The data acquisition system must be able to read out the camera at maximum speed (3500 Hz full frame 320x256) and be able to stream the data to disk continuously.  The realtime system needs to be able to calculate fringe phase estimates with $<$50ms latency for group delay tracking and $<$5mas for phase tracking.  This requires close coordination between the systems, including in data saving and using all fringe weights from both systems for fringe tracking.  The code will be the same for both systems, except for the fringe estimators for the 4-beam integrated optics combiner. More details on the MIRC-X system that is already on-sky can be found elsewhere in these proceedings (Kraus et al, Anugu et al.).

\subsection{Architecture}

The MIRC-X and MYSTIC software follows the CHARA standard ``client/server'' model and runs under Linux. The hardware such as camera, motors/actuators and sensors are controlled by C-written servers. A server is a program that accepts requests, performs a task and returns some data. The servers can then have multiple clients (GUIs, or other servers) that make requests to them.  All the MIRC-X servers run on the MIRC-X computer. All the MYSTIC servers run on the MYSTIC computer. Many other servers run on different machine inside CHARA, and are accessible via the standard mechanism (e.g shutter server, DL server...).

\begin{figure} [ht]
   \begin{center}
   \begin{tabular}{c} 
   \includegraphics[width=6in]{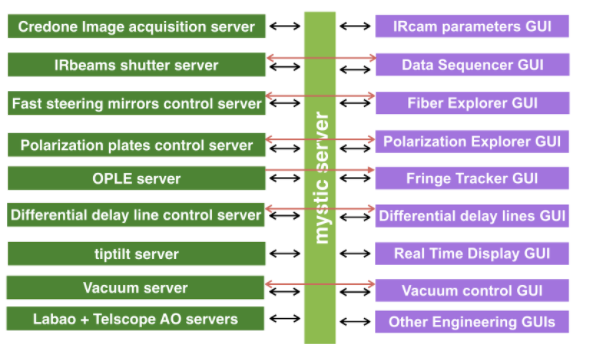}
  \end{tabular}
   \end{center}
   \caption[example] 
   { \label{fig:software21} 
The MYSTIC server communication with the various servers and GUIs.  The arrow marks indicate the communication.
}
\end{figure}

The GUIs clients are GTK based (Gimp Tool Kit). GTK is a software library for developing X windows programs. The GUI can be running on various machines, locally or remotely. They connect to the servers via the standard TCP/IP-socket-based CHARA protocols. Figure~\ref{fig:software21} presents the list of servers and clients that will be used to operate the MYSTIC instrument. The architecture is very similar for MIRC-X, the FSM server being replaced by the fiber server.

\subsection{C-RED One Data Acquisition System}

\begin{figure} [ht]
   \begin{center}
   \begin{tabular}{c} 
   \includegraphics[width=6in]{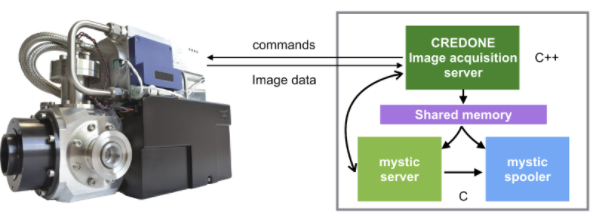}
  \end{tabular}
   \end{center}
   \caption[example] 
   { \label{fig:daq22} 
MYSTIC image acquisition software.  The C-RED One camera images are acquired to the MYSTIC computer and written to a shared memory. The mystic server reads the shared memory and do the image processing and fringe tracking. The C-RED One camera parameters  can be changed using GUIs connected to the mystic server. The mystic spooler reads the images from the shared memory and writes into fits data.  
}
\end{figure}

MIRC-X and MYSTIC each uses a C-RED One camera, an ultra-fast, ultra low read noise infrared camera based on the e-APD technology from First Light Imaging. Figure~\ref{fig:daq22} describes the data acquisition system. Two camera link cables (2 female SDR-26) are used to connect the camera with the data acquisition computer. There is a low-speed  serial connection for altering camera configuration in addition to the framegrabbing functionality.

The camera (inside the beam combiner lab) and the data acquisition computer (placed outside the beam combiner lab) are separated by a few tens of meters and for this reason, the camera-fiber-link extender system will be used (as already implemented for MIRC-X, with camera-fiber-link hardware from thinklogical). It uses a camera side unit and a frame grabber or computer side unit that are connected by multimode fiber optic cables, allowing camera link video support up to 350 meters from the host computer with no loss of signal and without the use of amplifiers. 

We use the Matrox Radient eV-CL frame grabber as recommended by First Light Imaging. It is installed on the PCIe 2.0 x 16 slot at the instrument computer. It allows data speed of a peak bandwidth of up to 4GB/s. This speed is enough for the C-RED One camera frame grabbing as it produces 540 MB/s data (320 x 256 pixels$^2$  x 16-bit x 3500 frames/s). 
The camera images are handled by the C-RED One image acquisition server. This program grabs the images from the camera to the MYSTIC computer and writes them to a circular-buffer based shared memory. The code is written in Qt 5.6 and C++.

The MYSTIC server runs on the same computer as the image acquisition server and reads the images from the shared memory and does pre-processing, power spectrum computation and fringe tracking. The MYSTIC server communicates with the C-RED One image acquisition server to modify the camera parameters.

The MYSTIC spooler is the data saving process. It reads images from the shared memory and writes to FITS files. It writes the fits header by reading parameters from the CHARA servers (OPLE, GPS, tip-tilt and AO servers). This process can be given low linux priority to allow the realtime processes (i.e., filling of circular buffers, fringe phase calculations) to have low latency.

\begin{figure} [ht]
   \begin{center}
   \begin{tabular}{c} 
   \includegraphics[width=6in]{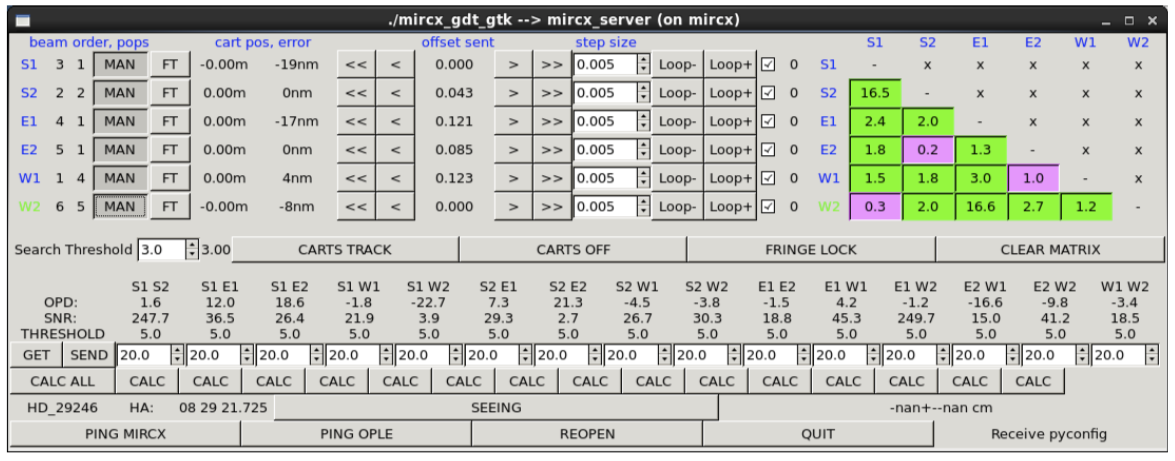}
  \end{tabular}
   \end{center}
   \caption[example] 
   { \label{fig:guis23} 
The GUIs for MIRC-X and MYSTIC will be the same and newly-written in gtk, although based on the python guis used originally by MIRC.  Here is an example of the new Group Delay Tracking GUI. The Loop-/Loop+ buttons are used to search for the fringes. Once the fringes are found the matrix buttons (in green color) are used to lock the fringes.
}
\end{figure}

\subsection{Example GUI}
We highlight one of our GUI interfaces, the Group Delay Tracker (GDT) GUI, currently in use for MIRC-X at CHARA.  During observing or fringe searching, the mircx server computes the SNR and group delay for all fringes in each incoming frame, accounting for the number of coherent and incoherent integrations specified by the user. The fringe loop (e.g search) is initiated by moving the CHARA delay lines in regular steps until the SNR reaches the required threshold. Then fringe are tracked by sending the measured OPD to the CHARA delay lines. The algorithm uses the closing triangles to lock fringes on more baselines. The tracking algorithm runs entirely on the mircx server and the user interacts with it via a dedicated GUI. Figure~\ref{fig:guis23} presents the group delay tracking GUI for MIRC-X.

\subsection{Software status}
We have a working set of GUIs along with a low-latency DAQ for MIRC-X that we are cloning for MYSTIC.  One small difference is we have chosen a server computer with more cores (and slightly lower CPU speed) in order to experiment with more computer-intensive realtime processing of the raw data frames, e.g. median filtering.

Once MIRC-X and MYSTIC are running simultaneously, we will need a more sophisticated fringe tracking algorithm that make use of all the fringe data from all both instruments simultaneously. This will also open up the possibility of phase tracking for bright enough targets.  Our team will be working on these algorithms over the next year to be ready in 2019 when both systems will be operating on-sky.

\section{Status and future work}
The MYSTIC project had a design review in 2018 March and received practical and useful feedback from the review panel. Importantly, the project received a green light to continue toward construction and the Michigan team is now ordering the custom optics and pieces needed.  

Table~\ref{tab:schedule} lists our expected schedule of remaining major milestones.

\begin{table}[ht]
\caption{MYSTIC Schedule (estimate) } 
\label{tab:schedule}
\begin{center}       
\begin{tabular}{|l|l|} 
\hline
\rule[-1ex]{0pt}{3.5ex} Period & Milestones   \\
\hline
2018 Summer & Build MIRC-X in Michigan Lab\\
 & Prototype new photometric channel architecture\\
 & Glue lenslet arrays to v-grooves/IO components \\
 & Finalize design of cryostat following prototyping\\
\hline
2018 Fall & Order cryostat \\
  & Build warm version of MYSTIC in lab \\
\hline
2019 Winter & Receive cryostat and integrate MYSTIC inside \\
& Alignment and end-to-end tests \\
\hline 
2019 Spring & Pre-ship review \\
\hline
2019 Summer & On-sky Commissioning \\
\hline
\end{tabular}
\end{center}
\end{table}

\acknowledgments 
MYSTIC is funded by the USA National Science Foundation (PI: Monnier, NSF-ATI 1506540) while the MIRC-X project is funded by the European Research Council (PI: Kraus, ERC, Grant \# 639889). We thank Nicolas Besacier (IPAG) for his work to measure the transmission and bending losses for a number of candidate fibers for MYSTIC and MIRC-X.

\bibliography{report} 
\bibliographystyle{spiebib} 

\end{document}